# Exploring the Panorama of Anxiety Levels: A Multi-Scenario Study Based on Human-Centric Anxiety Level Detection and Personalized Guidance

Longdi Xian [a*], Junhao Xu [b]

[a] Department of Psychiatry, Faculty of Medicine, The Chinese University of Hong Kong, Hong Kong

[b] Faculty of Computer Science and Information Technology, University of Malaya, Malaysia

**Abstract**

More and more people are under pressure from work, life and education. Under these pressures, people will develop an anxious state of mind, or even the initial symptoms of suicide. With the advancement of artificial intelligence technology, large language modeling is currently one of the hottest technologies. It is often used for detecting psychological disorders, however, the current study only gives the categorization result, but does not give an interpretable description of what led to this categorization result. Based on all these immature studies, this study adopts a person-centered perspective and focuses on GPT-generated multi-scenario simulated conversations. These simulated conversations were selected as data samples for the study. Various transformer-based encoder models were utilized in the study in order to integrate a classification model capable of identifying different anxiety levels. In addition, a knowledge base focusing on anxiety was constructed in this study using Langchain and GPT4. When analyzing the classification results, this knowledge base was able to provide explanations and reasons that were most relevant to the interlocutor's anxiety situation. The study shows that the developed model achieves more than 94% accuracy in categorical prediction and that the advice provided is highly personalized.

Anxiety disorder, Transformer, Nature Language Processing, Large Language Models

# 1 Introduction

Mental health is defined as a state of well-being on the mental, emotional, and social levels [8, 16, 34]. Abnormal anxiety is a very important factor that leads to mental health [3, 19, 43]. It is a kind of irritability caused by excessive worry about the safety of the life of the loved ones or oneself, the future destiny, and so on. Anxiety can sometimes lead to a depressed state of mind in people serious and even lead to symptoms of suicide [7, 20]. So it is very important to prevent anxiety in our daily life [42].

Mental health will benefit less from digital health interventions due to the complexity of life circumstances, and more research is needed to consider populations with a range of psychosocial complexities [5].

This study presents an innovative framework aimed at assessing and understanding human anxiety levels through a large-scale language model (LLM)-based approach. The core technology is a simulated conversation generated using the latest GPT-4.0 model, which covers multiple scenarios including education, daily life, work, and socialization, thus providing a rich data sample for the study. The framework incorporates a series of advanced transformer-based encoder models, including Bert [12], RoBERTa [11], DistilBERT [24], and ELECTRA [6]. The integration of these models allows the framework to accurately recognize different levels of anxiety states, such as no anxiety, mild anxiety, and severe anxiety. In addition, by combining Langchain and GPT4, a knowledge base focusing on anxiety was constructed by utilizing data on the components of anxiety, as well as anxiety-related simulation cases. This knowledge base can be called upon to obtain the most relevant explanations and causes of the interlocutor's anxiety condition, while also connecting to GPT4.0's ability to give the best individually tailored and reasonable advice. This study can be widely used in various fields such as mental health assistance, educational counseling, workplace stress management, and social skills training. By accurately recognizing and understanding an individual's anxiety state, more targeted support and intervention can be provided to help people better manage and reduce anxiety and improve their quality of life. The research contributions of this paper are as follows:

- This study utilized GPT-4, functioning as a Large Language Model (LLM), to generate an anxiety conversation dataset from a multi-scenario perspective.
- Trained and evaluated a human-centric, scene-based anxiety level prediction comprehensive model based on LLM, which can predict anxiety levels in different scenarios and provide reasonable explanations and suggestions.
- The latest GPT-4 as well as Langchain have been used for intelligent suggestions based on the individual's own session as well as intelligent session analysis. This enables the AI to interact with humans and automatically arrive at the best possible outcome for the person.

## 2 Literature Review

Although numerous national and organizational level approaches have taken to improve their mental health, students, workers, social peoples and others still suffer from high rates of mental health problems [13]. The COVID-19 pandemic is impacting mental health, but it is not clear how people with different types of mental health problems were differentially impacted as the initial wave of cases hit. The subject of [18] is to leverage natural language processing (NLP) with the goal of characterizing changes in 15 of the world's largest mental health support groups found on the website Reddit, along with 11 non–mental health groups during the initial stage of the pandemic. [10] apply natural language processing (NLP) to unstructured therapy transcript data from patients seeking treatment during the height of the pandemic in the United States between March 1, 2020 and June 9, 2020 to identify words associated with COVID-19 mentions. These terms can be categorized into 24 symptoms beyond the scope of diagnostic criteria for anxiety or depression. [21] hypothesize that using natural language processing (NLP) to explore social media would help to identify the mental health condition of the population experiencing insomnia after the outbreak of COVID-19. In this study, [36] discuss the application of NLP in psychotherapy and also a general analysis of existing systems was performed by comparing the responses given by the chatbot against a set of predefined user inputs pertaining to queries related to wellbeing and mental health. The general methodology involved in the creation of such chatbots includes the underlying NLP techniques like word embeddings, sentiment analysis, Models like Sequence-to-Sequence model and attention mechanism [36]. By integrating mental health assessment tools into the chatbot interface, along with regular therapy it can help patients deal with mild anxiety and depression. The research conducted by [22], as you described, appears to focus on using AI for "soft intelligence" analysis, particularly through examining UK tweets related to mental health during the COVID-19 pandemic. A series of keyword filters were used to clean the initial data retrieved and also set up to track specific mental health problems. Currently, there is no established methodology for distinctly characterizing the mental health impact of the pandemic, separate from pre-COVID-19 levels. Crucially, it is recommended to monitor a broader range of symptoms beyond those for anxiety and depression to comprehensively understand the full scope of COVID-19's effects on mental health [9]. The purpose of [23] was to explore the value of soft intelligence analyzed using NLP. Using an NLP platform, were able to rapidly mine and analyze emerging health-related insights from UK tweets into how the pandemic may be impacting people's mental health and well-being.

Mental health (anxiety, depressive symptom, and self-esteem) was assessed by self-administered questionnaires [17]. [26] aim to examine the relationship between psychosocial stressors and symptoms of depression and anxiety and the extent to which emotional support or resilient coping moderates the relationship between psychosocial stressors and maternal mental health during the first wave of the COVID pandemic. Strategies focused on bolstering coping and social support may be insufficient to improve maternal mental health during acute public health emergencies. [16] attempt to express and ontologize the relationships between different mental disorders and physical organs from the perspective of TCM, so as to bridge the gap between the unique terminology used in TCM and a medical professional. The results demonstrate that the proposed framework integrates NLP and data visualization, enabling clinicians to gain insights into mental health, in addition to biomedicine. [28] investigate potential protective and risk factors such as sociodemographics and psychological factors such as adaptation/coping. This knowledge may inform public health agencies on how to promote mental health in similar situations in the future. The COVID-19 pandemic has had wide-ranging impacts on mental health. However, less is known about predictors of mental health outcomes among adults who have experienced a COVID-19 diagnosis. [37] examines the intersection of demographic, economic, and illness-related predictors of depressive and anxiety symptoms within a population-based sample of adults diagnosed with COVID-19 in the U.S. state of Michigan early in the pandemic. Using cross-sectional data from the observational Hispanic Community Health Study, [29] examined the relationship between MetS and depression and anxiety in addition to testing moderating effects of gender and Hispanic heritage subgroups. Gender moderated the relationship between mental health and MetS, with women having a significant increase in the probability of MetS with depression ($p < .001$), anxiety ($p < .001$), or both ($p < .001$). [30] aim to estimate the prevalence of mental disorders and their associated factors among the general population in Korea. Future studies on this topic and efforts to increase the mental health treatment rate at a national level are needed. Second, examine the role of perceived social support in the association between SGM-ACEs and adult mental health. [5] study sexual and gender minority adverse childhood experiences (sgm-aces), perceived social support, and adult mental health. In total, 1819 self-identified SGM Texans completed an online survey inquiring about ACEs, SGM-ACEs, mental health, and demographic characteristics. [1] assess the associations between maternal mental health and dental anxiety level, dental caries experience, oral hygiene, and gingival status among 6- to 12-year-old children in Nigeria. Maternal mental health risk and depression do not seem to be risk factors for schoolchildren's oral health in

Nigeria.

# 3 Research Methodology

## 3.1 Research Design

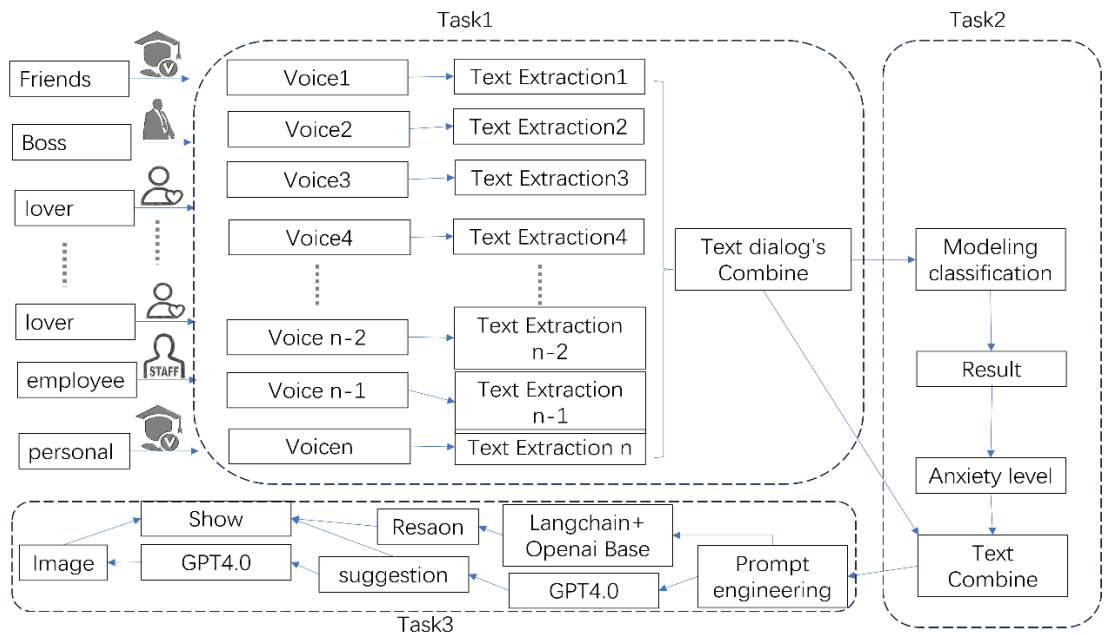

*Figure 1 Research Design*

  The framework (Figure 1) of this study consists of three main parts, task1, task2 and task3. task1 is a collection of sessions of people communicating in different contexts such as education, work, daily life and socialization, storing the conversion of text and storing the text of the sessions. The collection process of the dataset is shown in detail in 3.2.1. After the storage process is completed, the next task is task2, in which the final data are entered into the model for the first prediction of anxiety levels (see 3.2.2 Model Construction for details), and the predicted anxiety levels are combined with the final raw data in order to prepare for task3. In the next step of task3, the main purpose is to first combine the combined text using the cue engineering model to form a reason for calling the anxiety base constructed by langchain to obtain an individual-based judgment of the anxiety level (see 3.3 for details), as well as calling the cue text of the Artificial Intelligence Interface (GPT4.0) to obtain the person-centered advice for oneself. After obtaining reasonable suggestions, the multimodality of GPT4.0 will be called again to obtain the best and most vivid image in which all reasonable suggestions will be visualized. At the same time, the cueing project is called again to read the text base of anxiety built using Langchain+GPT4.0 to obtain the reason for this prediction (i.e., the sentence in the session that embodies the anxiety).

## 3.2 Dataset and Modelling

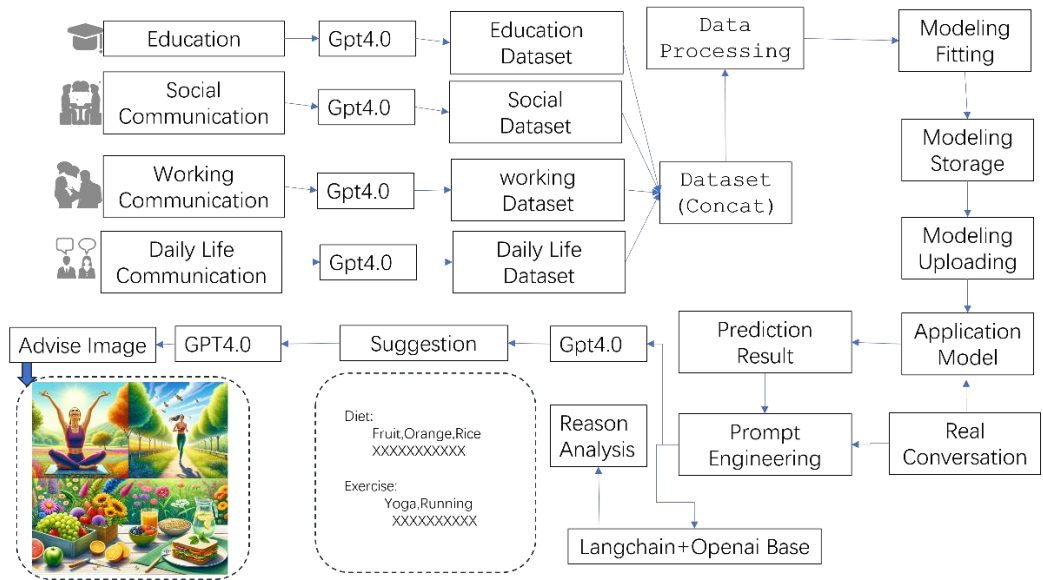

*Figure 2 Model Construction and Prediction of Results*

### 3.2.1 Dataset Generating

Based on the dataset shown in (Figure 2), the dataset is generated by using GPT-4.0 to generate simulated session data for four different scenarios (daily life, education, socialization, and work) through API connection combined with prompt engineering technique. The specific steps are as follows:

1.define the scene-specific prompt template.

$$SSC = LLM(Prompt) \qquad (1)$$

In this function, SSC (Scene-Specific Content) is the output generated by the function. SSC refers to the content that is specifically tailored to the scene described in the prompt. $LLM(\text{prompt})$ of the function represents the operation of a Large Language Model (GPT-4) when provided with a prompt. The $LLM$ takes the input prompt, processes it, and generates output text of different scenarios.

*Table 1 Prompts in Different Scenarios*

| Scenario | Prompt |
|---|---|
| Education | In educational scenarios, Chinese is used to generate conversations between pairs (teacher and student, parent and student, classmates, etc.) with different anxiety levels (no anxiety, a little anxiety, super anxiety). |
| Working | In a work scenario, Chinese is used to generate conversations between pairs of people (bosses and employees, coworkers, etc.) with different anxiety levels (no anxiety, a little anxiety, super anxiety). |
| Daily Life | In everyday life scenarios, Chinese is used to generate conversations between pairs (friends, brothers, couples, etc.) with different anxiety levels (no anxiety, a little anxiety, super anxiety). |
| Social | In social scenarios, Chinese is used to generate conversations between pairs of people (on blind dates, when docking clients, etc.) with different anxiety levels (no anxiety, a little anxiety, super anxiety). |

2.Call GPT-4.0 via API to input templates for each scenario.
• The chat function takes a list of prompts as input. It enters a loop that sends these prompts to the OpenAI GPT-4.0 model. The API call is configured to limit the length of the response to 150 tokens, which helps manage the response time and avoid timeouts. If the API call fails for any reason (e.g., a network error), the function catches an exception, waits 1 second, and then retries the request.

• Initialization: Initializes two empty lists (questions and answers) for storing generated questions and corresponding answers.

• Generate Questions and Answers:In each iteration (421 iterations are set for each scenario to obtain 421 pieces of data), it constructs a prompt (shown in Table 1) by adding instructions to the text to generate a Chinese dialog that reflects a little anxiety.

• The chat function is then called, and the results are captured. Each result (generated dialog) is added to the list of questions, and the corresponding anxiety level (anxiety level) is added to the list of answers.

3.Collect the Dialog Data Generated by GPT-4.0
For each scenario, the structure of the dataset is defined as follows:

• text column: conversation data in different scenarios.
• anxiety column: anxiety level (no anxiety, mild anxiety, significant anxiety).

The generated datasets are saved as daily.csv, education.csv, social.csv, working.csv respectively.

The above four datasets are merged into one dataset anxiety.csv. The data processing involved in the merging process can be represented by the following equation:

$$S_i = [D_i^{daily}, D_i^{education}, D_i^{social}, D_i^{working}], \qquad (2)$$

where $D_i^{daily}, D_i^{education}, D_i^{social}, D_i^{working}$ represent the datasets of daily life, education, social, and working scenarios respectively.

This formula is a data merging process that combines four different datasets (i.e., datasets for everyday, educational, social, and work scenarios) into a new dataset. Each dataset is stored as a CSV file containing information related to that scenario. In this process, using a new variable named $S_i$ to represent all the information in the new dataset.

### 3.2.2 Modelling Building

Overview of the model BERT (Bidirectional Encoder Representations from Transformers)[27]: As a breakthrough model, BERT introduces bi-directional pre-training for a more comprehensive understanding of linguistic context [25]. It has set new performance standards on many NLP tasks and has therefore become a focus of research and applications.
*RoBERTa (Robustly Optimized BERT Approach)*: RoBERTa is an improved version of BERT that significantly improves performance through longer training times, larger datasets, and finer tuning. Studying RoBERTa helps to understand how to optimize model performance by tuning the training process [15]. *DistilBERT*: As a "lightweight" version of BERT, DistilBERT aims to maintain similar performance as BERT while significantly reducing the size and complexity of the model. It is important for resource-limited application scenarios, and also demonstrates the possibilities of model compression and optimization [38]. *ELECTRA (Efficiently Learning an Encoder that Classifies Token Replacements Accurately)*: ELECTRA employs a different pre-training method, by distinguishing real words from other words. This approach shows better performance than BERT on some tasks, providing an interesting case study of the impact of different pretraining strategies [42]. BERT, RoBERTa, DistilBERT, and ELECTRA are four popular Natural Language Processing (NLP) models that are widely used for text prediction and other linguistic tasks. Meanwhile, the Transformer model is the basis for building BERT, RoBERTa, DistilBERT, and ELECTRA. The Transformer contains an Encoder and a Decoder, each part of which consists of multiple identical layers stacked on top of each other [27]. Each layer of the Encoder includes the Multi-Head Attention mechanism and a feed-forward neural network. The input is first processed by Word Embedding and Position Embedding. The decoder includes an attention mechanism to the encoder output in addition to a similar structure to the encoder. In BERT, RoBERTa, DistilBERT, and ELECTRA, the encoder structure is mainly used.

Construction of the model

**Word EmbeddingWord Embedding**

For the processing of text data, especially in deep learning models, Word Embedding is a key step. It converts the raw text into a numerical form that the model can understand and process. Below is a detailed description of the word embedding process [40], using the Chinese text provided as an example. First of all, the Chinese text needs to be segmented. Segmentation is the process of dividing continuous text into meaningful segments (words or characters), adding [CLS] tags to the beginning of each text and [SEP] tags to the end of each sentence [2]. For example, for text:

$T_{\text{text}}$= ['Person A: Hello,what are your plans for the weekend?Person B:Oh,nothing much']

$T_{\text{participle}}$= [[CLS], "Hello", ",", "weekend", "plans", "do", "what", "?", "Oh", ",", "nothing", "much", [SEP]]

Build a vocabulary list containing all occurrences of the word and its index.

Example: Glossary = {"Hello": 1, "weekend": 2, "plans": 3, ..., "not bad": N}

Next, the split text is converted into word vectors. This usually involves mapping each word to a fixed-size vector of real numbers. Assuming we use a pre-trained word embedding model (e.g., Word2Vec, GloVe, or BERT's Embedding layer), each word is converted to a D-dimensional vector.

Word vector: 'hello' → [0.25, -0.41, ... , 1.25]  (D = 100 )( is assumed in the example)

After converting all words to vectors, the original text is converted into a sequence of vectors. Each sentence may have a different length, and to deal with this, Padding or Truncating is usually required to ensure consistency of the input length. Sequence vector: [[0.25, -0.41, ... , 1.25], [...] , ... , [...]] Finally, these sequence vectors serve as inputs to the model. In your example, if the Label indicates the anxiety level, then the text is input into the model along with the label for training or prediction. Before word embedding: this column shows the result after the original text has been processed by the word segmentation. After word embedding: each word is converted into a vector (represented in this table by the simplifications E1, E2, E3... etc. simplified representation). In fact, each E is a D-dimensional vector (D is usually very large, e.g. 128, 256 or

higher)

*Table 2 Word Embedding Sample Table*

| ordinal number | Text | Label | Before Word Embedding (Segmentation Results) | After Word Embedding (Vector Representation) |
|---|---|---|---|---|
| 1 | Person A: Hello, what are your plans for the weekend? Person B: Oh, nothing special planned, might go to the park for a walk. Person A: Sounds good. | 0 (no anxiety) | [CLS], Person A, :, Hello, ,, weekend, plans, to do, what, ?, Person B, :, Oh, ,, nothing, special, planned, ,, might, go, to, the park, for a walk, ., Person A, :, Sounds, good, ., [SEP] | [Emb_CLS,E1, E2, E3, E4, E5, E6, E7, E8, E9, E10, E11, E12, E13, E14, E15, Emb_SEP] |
| 2 | Person A: We need to speed up the progress of this project. Person B: I know, I've been working overtime at night recently, hoping to catch up with the progress. | 1 (mild anxiety) | [CLS], Person A, :, We, need, to, speed up, the, progress, of, this, project, ., Person B, :, I, know, ,, I've, been, working, overtime, at night, recently, ,, hoping, to, catch up, with, the, progress, ., [SEP] | [Emb_CLS,E16, E17, E18, E19, E20, E21, E22, E23, E24, E25, E26, E27, E28, E29, E30, Emb_SEP] |
| 3 | Person A: Have you been sleeping well lately? Person B: Actually, not really. I've been thinking about our loan issue. | 1 (mild anxiety) | [CLS], Person A, :, Have, you, been, sleeping, well, lately, ?, Person B, :, Actually, ,, not, really, ., I've, been, thinking, about, our, loan, issue, ., [SEP] | [Emb_CLS,E31, E32, E33, E34, E35, E36, E37, E38, E39, E40, E41, E42, E43, E44, Emb_SEP] |
| 4 | Person A: What do you think of the new colleague? Person B: They're very nice, I think we can cooperate well and have a pleasant working relationship. | 0 (no anxiety) | [CLS], Person A, :, What, do, you, think, of, the, new, colleague, ?, Person B, :, They're, very, nice, ,, I, think, we, can, cooperate, well, and, have, a, pleasant, working, relationship, ., [SEP] | [Emb_CLS,E45, E46, E47, E48, E49, E50, E51, E52, E53, E54, E55, E56, E57, E58, E59, Emb_SEP] |
| 5 | Person A: How are your preparations for the new project going? Person B: There's still a lot of work to be done, and the pressure is high. | 1 (mild anxiety) | [CLS], Person A, :, How, are, your, preparations, for, the, new, project, going, ?, Person B, :, There's, still, a, lot, of, work, to be done, ,, and, the, pressure, is, high, ., [SEP] | [Emb_CLS,E60, E61, E62, E63, E64, E65, E66, E67, E68, E69, E70, E71, E72, E73, Emb_SEP] |

**Position embedding and Final embedding**

1.Use the first participle data in the word embedding as an example(Table first row):

", "Person A", ":", "Hello", ",", "weekend", "plans", "to do", "what", "?", "Person B", ":", "Oh", ",", "nothing", "special", "planned", ",", "might", "go", "to", "the park", "for a walk", ".", "Person A", ":", "Sounds", "good", ".", " [SEP] "]

Generate positional numbers: Each word is assigned a positional number. The position numbers start from 0 and increase in order.

Position number: [0, 1, 2, 3, 4, 5, 6, 7, 8, 9, 10, 11, 12, 13, 14, 15, 16, 17, 18, 19, 20, 21, 22, 23, 24, 25, 26, 27, 28, 29,30,31]

2. Position embedding matrix

In the Transformer model, a positional embedding matrix is usually predefined, where each row corresponds to a positional embedding vector. For example, if we set the maximum sentence length to be 32 (in the above example) and the embedding dimension to be D, then the size of the positional embedding matrix will be 32×D.

3. Add positional embedding vectors to word embedding vectors

The word embedding vector of each word will be summed with its corresponding positional embedding vector to obtain the final embedding vector that contains both lexical content and positional information.Sum the location embedding vectors: for each word, its

Final embedding vector= word embedding vector+location embedding vector

Word embedding vector: Suppose our word embedding vector is $E_i$, where $E_i$ denotes the embedding vector of the word in the sequence[40]. Positional embedding vector: Let the positional embedding matrix be $P_2$, where $P_i$ denotes the positional embedding vector of the position in the sequence aaa[39].For example, if the word embedding vector of "你好" is $E_1$, and its position number is 1, then the final embedding vector is $E_1 + P_1$, where $P_1$ is the embedding vector of position 1. The detailed process is shown below:

WordEmbedding:

$$E_0, E_1, E_2, E_3, \dots$$

PositionEmbedding:

$P_0$ (correspondingto[CLS])
$P_1$ (correspondingto"PersonA",")
$P_2$ (correspondingto"：")
$P_3$ (correspondingto"Hello")
…

*Table 3 Final Embedding Vector Process Table*

| Segmentation results | Word embedding vector (example) | Position number | Position embedding vector (example) | Final embedding vector (example) |
|---|---|---|---|---|
| [CLS] | [0.1, 0.0, 0.0] | 0 | [0.0, 0.0, 0.0] | [0.1, 0.0, 0.0] |
| Person A | [0.1, 0.2, 0.3] | 1 | [0.1, 0.1, 0.1] | [0.1, 0.2, 0.3] |
| ： | [0.1, 0.3, 0.2] | 2 | [0.2, 0.2, 0.2] | [0.2, 0.4, 0.3] |
| Hello | [0.2, 0.2, 0.4] | 3 | [0.3, 0.3, 0.3] | [0.4, 0.4, 0.6] |
| , | [0.1, 0.1, 0.1] | 4 | [0.4, 0.4, 0.4] | [0.4, 0.4, 0.4] |
| Weekend | [0.3, 0.2, 0.1] | 5 | [0.5, 0.5, 0.5] | [0.7, 0.6, 0.5] |
| … | …… | .. | …… | …… |
| …… | | | | |

Word embedding vectors: this column shows hypothetical word embedding vectors.

Position number: this column shows the position number of each word in the sentence.

Position embedding vectors: this column shows the hypothetical position embedding vectors corresponding to the position numbers. In practice, these vectors are usually generated by the positional embedding part of the model and have the same dimensions as the word embedding vectors.

Final Embedding Vector: This column shows the result of adding the word embedding vector and the position embedding vector, representing the final embedding vector used as input to the model.

**Multi-Head Attention and FFNN**

The Multi-Head Attention mechanism is a key component of Transformer models [32], such as BERT, Roberta, which have revolutionized the field of natural language processing. This mechanism allows the model to jointly attend to information from different representation subspaces at different positions [31].

Assume that the word embedding and positional embedding vector dimensions for each word are $d_{\text{model}}$, using $F_3$ above (i.e.: '你 好') as an example, $E_3$: [0.2,0.2,0.4] and $P_3$: [0.3,0.3,0.3]. Suppose we use 2 heads for the attention computation, so each head processes a vector of dimension $d_{\text{model}}/2$. Split the embedding vector for each word into 2 heads. For example, $E_{3,\text{head1}} = [0.2,0.2]$, $E_{3,\text{head2}} = [0.4,0.2]$.

For each head, we use scaled dot product attention computation. The attention calculation formula [41] is:

$$\text{Attention}(Q, K, V) = \text{softmax}\left(\frac{QK^T}{\sqrt{d_{\text{head}}}}\right)V, \quad (3)$$

Where $Q, K, V$ are Query, Key, Value and dk is the dimension of the key. Assuming $Q = K = V = E_{i,\text{head}}$ where $E_{i,\text{head}}$ is the $i$-th head of the embedding vector for the word, and $d_k$ is the dimension of the key. The attention output for each head is given by:

$$\text{head}_i = \text{Attention}(E_{i,\text{head}}, E_{i,\text{head}}, E_{i,\text{head}}), \quad (4)$$

Here, the index $i$ of $V_i$ corresponds to the embedding vector index of each word.

The Feed-Forward Neural Network (FFNN) is used to further process the output of the self-attention mechanism [33]. The structure of this network can be formulated as follows:

1. First linear layer: This layer linearly transforms the output of the self-attention layer. If the dimension of the input is $d_{\text{model}}$, the dimension of the output of this layer may be $d_{\text{ff}}$ (e.g., in the original Transformer model, $d_{\text{model}} = 512$, $d_{\text{ff}} = 2048$).

2. Activation function, or ReLU (Rectified Linear Unit): It introduces nonlinearity and allows the network to learn more complex patterns. Assuming that our input is the output $X$ of the self-attention layer, the feedforward neural network can be represented as [14]:

$$\text{FFNN}(X) = \max(0, XW_1 + b_1)W_2 + b_2, \quad (5)$$

where $W_1$ and $W_2$ are the weight matrices of the network, and $b_1$ and $b_2$ are the bias terms. Here, $\max(0, z)$ is the ReLU activation function.

3. Second linear layer: This layer again transforms the data back to the original dimensional $d_{\text{model}}$:

$$X_2 = X_1 W_2 + b_2, \quad (6)$$

where $W_2$ and $b_2$ are the weights and biases of the second linear layer.

### Hierarchical Processing

The output of each layer becomes the input of the next layer, and this process is repeated in all layers of the encoder in the Transformer. As the data passes through each layer, the model gradually fuses and refines the information to capture the deeper semantic features of the text. After processing through multiple layers of the Transformer, the final output of the $[CLS]$ marker is a high-dimensional vector containing information about the entire sequence, denoted as $h_{[CLS]}$. The output of $[CLS]$ labeling is passed to a fully connected layer for classification [35].

$$\text{Output}_{\text{classification}} = \text{softmax}(W_c \cdot h_{[CLS]} + b_c), \tag{7}$$

Where, $W_c$ and $b_c$ are the weights and biases of the classification layer. Together, these steps accomplish the conversion from input text to classification results, enabling the model to achieve effective prediction (0, 1, 2) and classification (no anxiety, a little anxiety, particular anxiety) on different NLP tasks.

### Model intergrate

In order to make the results more rigorous and accurate, this study will use the method of integrating models by selecting the more optimal model for integration in different scenarios. The best results will be obtained by utilizing the intersection form. For example, if Bert, ELECTRAT, RoBERTa perform well in educational scenarios, then the results are denoted by:

$$T_{\text{Bert}} \cap T_{\text{ELECTRAT}} \cap T_{\text{RoBERTa}}, \tag{8}$$

Where ∩ represents the intersection symbol. When all three results are not the same, the prediction with the higher accuracy is chosen. If two of them are the same, then the result that is the same for both is chosen.

## 3.3 Anxiety Base

### 3.3.1 Building Anxiety's Base Steps

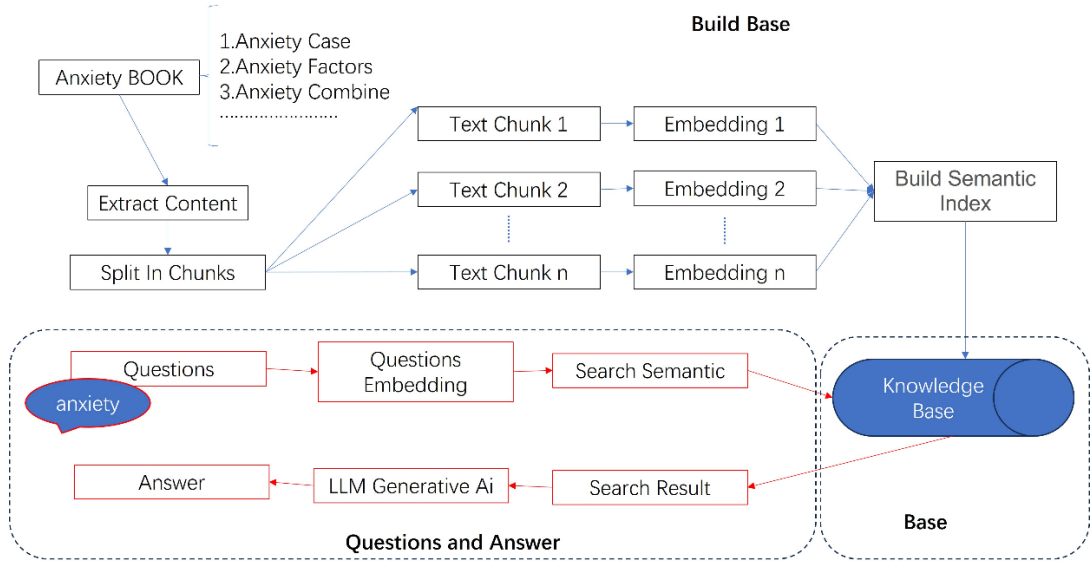

*Figure 3 Building Anxiety's Base Steps*

Figure 3 shows how to build an anxiety knowledge base based on human conversations using langchain and GPT-4.0. First, the data source is prepared for building the knowledge base by setting the GPT-4.0 API key and loading documents from a specified directory using DirectoryLoader. Next, the documents are segmented using RecursiveCharacterTextSplitter to break large documents into smaller chunks that are easy to process. Then, initialize OpenAIEmbeddings to generate embedding vectors for the document, which are key to converting the document into a numerical representation for machine learning tasks. Afterwards, a Chroma instance is created to store the document and its embedding vectors in a persistent catalog for quick retrieval and use. In addition, a system message template (system_template) is set up to provide contextual information to the user and to guide the way of answering. Initialize the message list and prompts to build the interaction logic of the Q&A system. Finally, use ConversationalRetrievalChain combined with langchain and GPT-4.0 models to create the Q&A system, which is able to process incoming questions and provide answers. This process involves converting from textual data to embedded vectors, which are then used to build a Q&A system capable of answering specific questions, especially for anxiety-related queries.

### 3.3.2 Vector DB

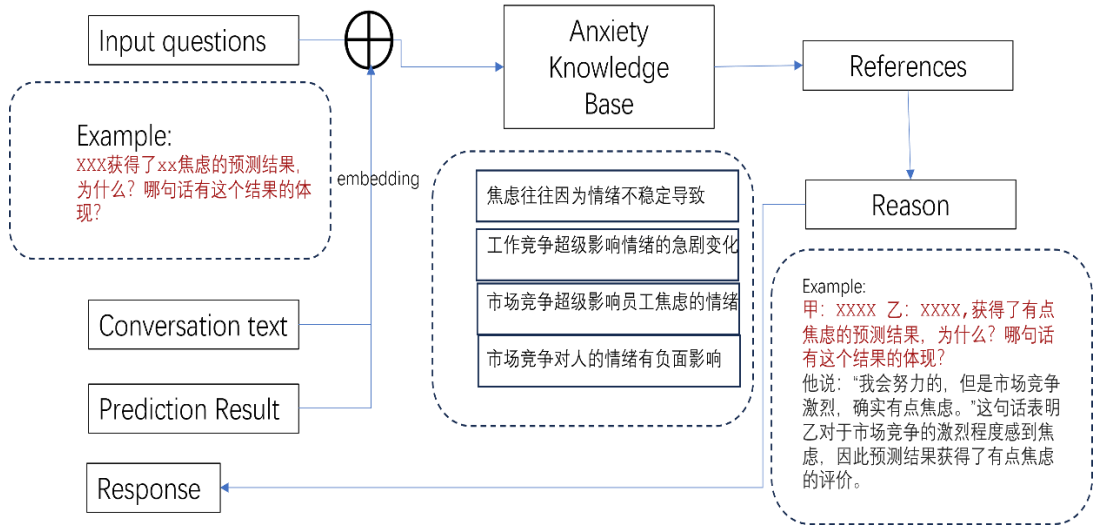

*Figure 4 Vector DB Invocation Process*

Figure 4 shows the specific anxiety base calling process. First, the prediction results of the anxiety levels obtained in different scenarios and all the dialog contents are embedded into the input question.

$$IQ = [C_{\text{dialog}}, C_{\text{Results}}], \tag{9}$$

Where $IQ$ is the input question, $C_{\text{dialog}}, C_{\text{Results}}$ denote the content of the session and the predicted results respectively.

For example:

"A: XXX? B: I will do my best, but the market competition is intense. I do feel a bit anxious... What is the predicted level of anxiety (xx)? Why? Which sentence reflects this result?"

The knowledge base constructed using langchain is then invoked and queried within the knowledge base to obtain the best reference information. After obtaining the reference information, the corresponding analysis of the reason for the anxiety level rating of the two-person conversation based on the scenario at that time is returned.

For example:

A: XXX? B: I will try my best... In the middle of the sentence, B says, "I will try my best, but the market competition is fierce, and I do feel a bit anxious."

This statement suggests that 'B' is anxious about the level of competition in the market, and therefore the prediction receives a somewhat anxious rating.

# 4 Experiments

## 4.1 Loss Functions and Optimizers

In order to make the model more accurate during the construction of the model, this experiment used 5 cross-validation and added a loss function and optimizer to each model construction. Sparse Categorical Crossentropy function was used [4].

$$L = -\sum_i \sum_j y_i \log(P_i), \qquad (10)$$

where $y_i$ is the true label and $P_i$ is the probability predicted by the model.

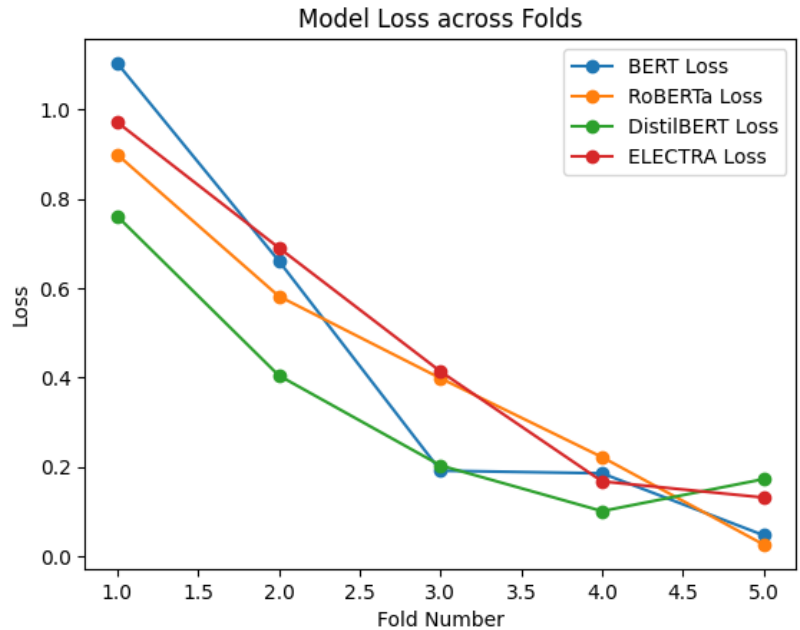

*Figure 5 Model Loss across Folds*

Figure 5 shows the Model Loss from the 5 cross-validations. All models show a general trend of decreasing loss across folds, suggesting improved performance in each fold. RoBERTa tends to have the lowest loss values in most folds, suggesting that it may be the most efficient model for that particular dataset. BERT and ELECTRA have similar patterns, but ELECTRA has higher losses in the first fold. They both improve significantly in subsequent folds. DistilBERT, a lighter version of BERT, performs very well, especially in the middle part of the fold, which is impressive considering its reduced size and complexity. At the same time, RoBERTa and DistilBERT seem to perform more consistently across folds than BERT and ELECTRA, which have larger initial losses. From these loss values alone, RoBERTa seems to be the most efficient model, combining low losses and consistency.

## 4.2 Hyperparameters and cross-validation

In this study, some hyperparameters are set and the model is trained and validated. The hyperparameters mainly include:

- **Learning Rate:** Set to $1e-5$ in the optimizer. Adam is a popular optimization

algorithm in machine learning and deep learning. It's known for its effectiveness in handling sparse gradients and its adaptiveness in different problems. The learning_rate=5e-5 sets the learning rate to 0.00005. The learning rate is a crucial hyperparameter that influences how much the model's weights are adjusted during training with respect to the loss gradient. A smaller learning rate means the model will learn more slowly, reducing the risk of overshooting the minimum of the loss function, but potentially taking more time to converge.

- **Batch Size:** Set to 8, this is the number of samples used in each training iteration to calculate the gradient and update the model weights.

- **Number of Training Cycles (Epochs):** Set to 2, meaning that the entire training set will be traversed twice.

- **Cross-Validation (StratifiedKFold):** Used in this study for cross-validation to ensure that each partition of the dataset is as similar as possible in terms of class distribution. $k$ is set to 5, meaning that the entire dataset is evenly divided into 5 parts. Each part is rotated as the validation set, and the rest serves as the training set, helping to reduce the bias and variance of the model and improve its generalization ability.

## 5 Result

### 5.1 Dataset

We simulate the conversation process of human scenarios and use the LLM of GPT-4 to generate a dataset (Anxiety Conversation Dataset under Different Scenario), which contains conversations between two people under four scenarios, namely, education, work, daily life, and socialization, and includes labels of anxiety levels. To make the scenarios more comprehensive and informative, all the scenarios include conversations between bosses and subordinates, friends, couples, strangers, coworkers, neighbors, and so on. At the same time, the conversation is in line with the logic of people's communication and also has a close relationship between the conversation. The basic structure of the dataset is shown in Table 4.

*Table 4 Anxiety Conversation Dataset under Different Scenario*

| Scenarios | 0/no anxiety | 1/a little anxiety | 2/particularly anxiety |
|---|---|---|---|
| Education | 154 | 278 | 231 |
| Social Communication | 221 | 298 | 125 |
| Daily Life Communication | 220 | 217 | 213 |
| Working Communication | 276 | 352 | 102 |

## 5.2 Modeling Performance

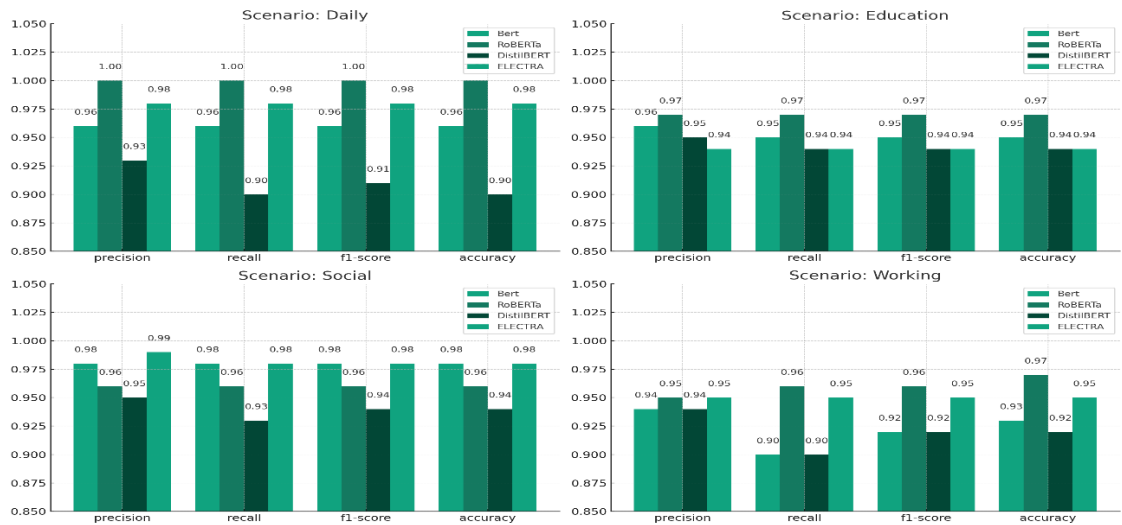

*Figure 6 Visualization of Classification Report for Anxiety Prediction in Different Scenarios*

Figure 6 shows the performance of different models for predicting anxiety levels in different scenarios, specifically Precision, Recall, F1-Score, and Accuracy. By analyzing each of these metrics, RoBERTa performs best in the "Daily" scenario, where it achieves a perfect score of 1.00 on all metrics, and very well in the "Social" scenario, where it achieves a perfect score of 1.00 on all metrics. The ELECTRA (0.98) model performs almost perfectly in this scenario, especially in accuracy (0.98) and precision (0.99). Bert and RoBERTa perform similarly in the "Education" scenario, where Bert (0.95) and RoBERTa (0.97) perform very well. The performance of the models in the "Work" scenario is more balanced, although RoBERTa (0.97) is slightly more accurate, the other models also perform well. Although the performance of the models varies across scenarios, RoBERTa generally performs the best, especially in the "Daily" (1.00) and "Education" scenarios (0.97). However, ELECTRA or other models may be more appropriate in certain situations, for example ELECTRA and Bert are more appropriate than RoBERTa in the 'Social' scenario.

In order to make the results of the study more rigorous and precise, this study will be conducted using the integration of the models. RoBERTa will be used in the scenarios of education, work, and daily life. ELECTRA and Bert will be used in social scenarios, and an intersection result will be obtained while using social scenarios, i.e., if $T_{\text{ELECTRA}} = T_{\text{Bert}}$, then the result is $T_{\text{ELECTRA}} \cap T_{\text{Bert}}$; if $T_{\text{ELECTRA}} \neq T_{\text{Bert}}$, then the result of $T_{\text{RoBERTa}}$ will be chosen, i.e., $T_{\text{RoBERTa}} \cap T_{\text{ELECTRA}} \cap T_{\text{Bert}}$. where ∩ means intersection.

## 5.3 Application

*Table 5 Application in Different Scenario Conversation*

| Scenario | Text | Result | Reason | Suggestion |
|---|---|---|---|---|
| Education | Student: Teacher, I'm a bit worried about this math problem. I tried my best, but I still feel like I don't understand it enough. Teacher: It's okay, anxiety is normal. Let's take it slowly and solve the problem step | A little Anxiety | The phrase "I'm a bit worried about this math problem. I tried my best, but I still feel like I don't understand it enough." | Take a break and relax outdoors as needed. |

| | by step. | | shows an anxious mindset. | |
| --- | --- | --- | --- | --- |
| Social | Individual: Hey, everyone! It's great to see you all. Tonight is going to be a lot of fun for sure. Friend: Welcome! We're also glad you could make it. Let's enjoy the evening together! | No Anxiety | The individual has a positive mindset and does not exhibit signs of anxiety. | Maintain the mindset and engage in appropriate physical activities. |
| Working | Employee: Manager, I'm concerned that I may have missed some details in this project report. Manager: No problem, we can review it together. The important thing is that you have already put in a lot of work, and we can improve it together. | A Little Anxiety | The employee shows a slight trace of anxiety with the statement "I'm concerned that I may have missed some details". | Take a break, shift attention, and engage in simultaneous communication. |
| Daily Life | Individual: Honey, I feel like everything in our household is under control, and everything is going smoothly. Family: That's great! It's wonderful to have you. We also think everything is going well. | No Anxiety | The individual has a positive mindset and does not exhibit signs of anxiety. | Continue to maintain a positive mindset and consider visiting entertainment venues like cinemas and amusement parks. |

Table 5 demonstrates the application of the proposed framework in multiple scenarios in practice, where the framework effectively recognizes emotional and affective states such as anxiety or anxiety in different conversations. This recognition capability is crucial for multiple scenarios such as education, work, and daily life, as it helps to better understand individual needs and reactions. At the same time, the framework not only recognizes emotional states but also provides personalized suggestions. This advice is based on the analysis of the content of the conversation, reflecting a highly customized and targeted application approach. For example, in the educational scenario, the relaxation suggestions made are specific to the student's anxiety. All of the resultant outputs that include the prediction of anxiety, the reasons given based on the results, and the individual suggestions given based on the results and the session achieved the desired results.

The framework not only demonstrates a powerful tool for analyzing conversations but also highlights the great potential for applying such tools to real-world situations. Through continued research and development, such frameworks can play an important role in helping individuals understand and manage their emotional and psychological states.

# 6 Conclusion and Future Work

The core outcome of this study is the development of a framework that can effectively recognize and manage anxiety. This framework demonstrated strong adaptability and accuracy in different scenarios such as education, socialization, work and daily life. In this study, the LLM in GPT4.0 was first utilized to generate a dataset including four scenarios of education, socialization, work and daily life (Anxiety Conversation Dataset under Different Scenario). This dataset has reached the standard of data samples after analysis and optimization. Meanwhile, RoBERTa, ELECTRA, DistilBERT and Bert were utilized to carry out the model construction based on this dataset. In terms of anxiety level prediction, different models showed significant differences in different scenarios. In particular, RoBERTa performed particularly well in the "daily" and "educational" scenarios, while ELECTRA was more effective in the "social" scenario. Model integration: The study demonstrates the effectiveness of model integration by combining RoBERTa, ELECTRA and Bert in different scenarios. The intersection of ELECTRA and Bert in the "social" scenario demonstrates that combining the predictions of multiple models in a given scenario can improve accuracy and reliability. The study focuses not only on identifying anxiety states, but also on providing individual-based cause analysis and providing sound person-centered advice. This study utilized Langchain+GPT4.0 to construct a systematic session function of the anxiety knowledge base, which was invoked through the hierarchical prediction results of anxiety were obtained to provide the expected causes. At the same time harm connection GPT4.0 obtained the best reasonable suggestions, these suggestions and interpretable reasons based on the conversation content analysis, reflecting a high degree of customization and relevance, which is of great value in helping individuals to understand and manage their emotional states. Although this study has achieved the desired results, due to factors such as dataset and scenario limitations, it is important to continue to explore and optimize model integration methods in different scenarios to improve the accuracy and efficiency of anxiety recognition. Extend the scope of the study to apply the framework to more diverse scenarios and populations, such as children, the elderly, or specific occupational groups. It remains important to develop real-time data processing capabilities for immediate anxiety monitoring and feedback to enhance the utility and interactivity of the framework.